\documentclass[a4paper,fleqn]{cas-sc}

\usepackage[authoryear,longnamesfirst]{natbib}
\usepackage{amsmath,siunitx}
\usepackage[referable]{threeparttablex}

\usepackage{setspace} 

\begin{document}
\let\WriteBookmarks\relax
\def\floatpagepagefraction{1}
\def\textpagefraction{.001}
\shorttitle{Skyglow Change after LED Lighting Retrofit}
\shortauthors{Hung et~al.}

\title [mode = title]{Changes in night sky brightness after a countywide LED retrofit} 

\author[1]{Li-Wei Hung}
\ead{li-wei_hung@nps.gov}
\author[1]{Sharolyn J Anderson}
\author[1]{Ashley Pipkin}
\author[1]{Kurt Fristrup}
\address[1]{Natural Sounds and Night Skies Division, National Park Service, 1201 Oakridge Dr., Suite 100, Fort Collins, CO 80525, USA}

\begin{abstract}
The US National Park Service (NPS) Night Skies Program measured changes in sky brightness resulting from a countywide lighting retrofit project. The retrofit took place in Chelan County, a gateway community to North Cascades National Park and Lake Chelan National Recreation Area in Washington State. The county retrofitted all 3,693 county-owned high pressure sodium (HPS) street lamps to full cutoff LEDs. This number is about 60\% of the County's total outdoor street and area lights. About 80\% of the newly installed lights were 3000K in color temperature and 20\% were 4000K. The 4000K LEDs were used to meet Washington State Department of Transportation guidelines. To measure sky brightness, we used the NPS night sky camera system before the retrofit started in 2018 and after its completion in 2019. These images were photometrically calibrated and mosaicked together to provide hemispherical images in $V$ band. For comparison with our ground-based measurement, we obtained the satellite imagery taken by Visible Infrared Imaging Radiometer Suite (VIIRS) onboard the Suomi National Polar-orbiting Partnership satellite. Our measurements show that the post-retrofit skyglow became brighter and extended higher in the sky, but upward radiance, as measured by the day-night band radiometer, decreased. These divergent results are likely explained by a substantial increase in light emitted at wavelengths shorter than 500 nm, and a relative decrease in zenith light emission due to better shielded luminaires. These results also demonstrate that earlier models relating VIIRS day-night band data to skyglow will – at a minimum – require substantial revision to account for the different characteristics of solid state luminaires.
\end{abstract}
 


\maketitle

\section{Introduction}
As of 2017, 19\% of outdoor lighting in the United States (US) had been converted to light-emitting diodes (LEDs), resulting in energy savings of 1.1 quadrillion British thermal units --- just over 1\% of US total energy consumption \citep{usdoe}. In addition to reducing energy consumption and extending service life, compact LED lamps can provide more precise control over the spectrum, light level, and projection of light than legacy lamps, such as High-Pressure Sodium (HPS) lamps, can. This transition in lighting will dramatically alter skyglow across the planet, but it is unclear if the change will improve or degrade the quality of the night sky. 

There is no consensus on whether current practices for LED lighting retrofits will increase or decrease skyglow. USDOE claims the increased directivity associated with LED upgrades will result in reduced skyglow \citep{kinz17}. While directivity and shielding are important, they are not the only factors at play. An LED retrofit often results in an increase of shorter wavelength light.  When lumens are equal the bluer spectral tone of most LEDs results in increased atmospheric scattering \citep{lugi14}. To accurately model skyglow, we must also identify how the shift to LEDs changes spectral power distribution (SPD) and luminous flux \citep{duri14}.  

Because of the reduced financial cost of lumens provided by LED technology, increases in the amount of artificial light is expected to grow across the planet \citep{kyba17, saun12, bore15}. \citet{lugi14} predicts the retrofit to LEDs from HPS could result in a night sky three times brighter, depending on the change in SPD. The USDOE, relying heavily on assumed implementation of responsible lighting practices such as mandatory shielding and dimming, states that residents near the city should not expect to see increased skyglow and that conditions may improve \citep{kinz17}. With so many conflicting statements managers making decisions about retrofits do not have consistent guidance to inform decisions.

Measurements of nighttime lighting have typically used the Day-Night Band (DNB) of the Visible Infrared Imaging Radiometer Suite (VIIRS) onboard the Suomi National Polar-orbiting Partnership satellite \citep{lee14}. VIIRS data are the dominant source of studying large scale outdoor lighting \citep{falc16, kyba17, bare18}. There are, however, two major concerns with using satellite data to characterize skyglow after an LED retrofit. First, satellite estimates of skyglow are uniformly derived from upward radiance, which can mischaracterize artificial light along the horizon. Improved shielding and increased atmospheric scattering from LEDs may exacerbate these inaccurate estimates \citep{lugi14}. Second, there are concerns about “blue blindness.” This refers to the inability of VIIRS to detect short wavelength light ($<$500nm), which has been cited in previous studies measuring light pollution during the shift to LEDs \citep{kyba17, bare18, falc16}. The measured radiance values can go down not because of decreases in light, but because VIIRS does not sense a significant portion of the radiation of these bluer lights.  Without accurate measurements, we may mischaracterize this rapid global change.  

Given these concerns, direct measurements of skyglow are crucial to overcoming the remote sensing limitations of the satellite data. There are currently no studies comparing VIIRS data with calibrated camera data that accurately measure skyglow from the horizon to the zenith.  

\begin{figure}
    \centering
    \includegraphics[width=0.7\textwidth]{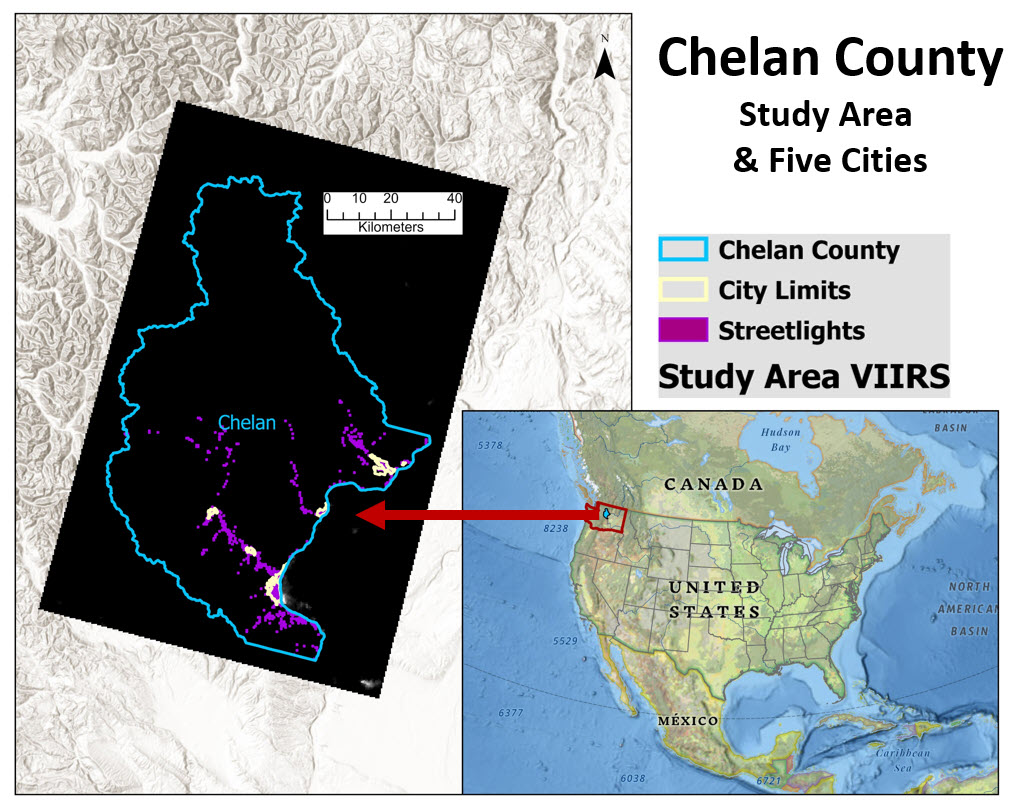}
    \caption{Location map for Chelan County Study Area with county outlined, five cities within the county and a overview map of United States with Washington State highlighted and Chelan County near the center of state. }
    \label{StudyArea}
\end{figure}

In 2018, Chelan County, Washington State, US (Figure~\ref{StudyArea}) began an LED retrofit that was completed in 2019.  This presented a perfect opportunity to measure skyglow after a direct one-for-one LED lighting retrofit. An important county objective for this retrofit was to conserve energy, but there was also interest in reducing the skyglow visible to many protected areas located within the county, including North Cascades National Park and Lake Chelan National Recreation Area. Community awareness about the negative impacts of light pollution and increased short wavelength light on ecology \citep{long04, holk10, manf17} and human health \citep{chep09, cho15, falc11}, as well as on recreation \citep{coll13} and local economies \citep{mitc19}, encouraged the county to request that information be collected before and after the retrofit to  determine if a reduction in skyglow was achieved. The county retrofitted 3,693 HPS drop lens luminaires to full cutoff LED luminaires.  

Color temperature measurements of typical HPS lamps range from 2100K-2500K. Most of the retrofitted LEDs were 3000K; 717 lights were retrofitted with 4000K LEDs to meet Washington State Department of Transportation guidelines \citep{wash19}. Crude estimates of around 18,400 lumens per light were calculated from wattage information of old HPS lighting. Estimates for new LED lighting, based on specifications provided by Chelan County, were approximately 9,500 lumens for each light.  

This paper uses two methods to analyze the skyglow brightness before and after an LED retrofit in Chelan County Washington. We measured the sky brightness directly using the National Park Service (NPS) night sky calibrated CCD camera system \citep{duri07}. This camera system provides high-resolution, panoramic measurements in the $V$ spectral band, closely matching the Johnson-Cousins $V$, from the horizon to the zenith.  We then compared the NPS CCD camera skyglow measurements with VIIRS measurements of upward radiance.

Due to the limited spectral sensitivity and confined observing angle, we believed that VIIRS data would underestimate skyglow after the one-for-one LED retrofit in Chelan County. Difference in the spectral characteristics of the relevant sources, filters, and detectors adds the complexity for this study. \citet{cao14} showed a nice comparison between the spectral curves of the VIIRS DNB, LED, and HPS. \citet{duri07} provided the details of the $V$ band filter and how it compares to the spectral response of the human eye. Using both satellite imagery and calibrated camera observations, we highlight the challenges of quantifying this change using only VIIRS satellite data. This case study provides insight on how LED lighting retrofits alter skyglow.  

\section{Methods}

\subsection{Satellite Measurements of Upward Radiance}

\subsubsection{Nighttime Imagery}
Satellite imagery is an efficient and consistent way to get a spatial perspective of a particular location, in this case before and after snapshots of Chelan County streetlights. The VIIRS instrument collects the highest spatial and temporal resolution nighttime imagery over the area of interest. This instrument has a DNB (day-night band) designed to collect high-quality radiometric data at night. The VIIRS DNB data are collected daily, with a local overpass at approximately 01:30 AM. VIIRS DNB has a spectral range from 500-900 nm, a 12-bit radiometric resolution, and spatial resolution of 742 m at nadir (resampled to 500 m for composite images).  

We downloaded six of the North American  monthly average composites (\textit{Tile  1\_75N180W}) that are freely available from Payne Institute at the Colorado School of Mines (April, May, and June of 2018 and 2019) \citep{elvi17} and selected the datasets that were stray light corrected by the Payne Institute \citep{mill13}.  We clipped the satellite images to a rectangular area encompassing Chelan County (Figure~\ref{StudyArea}).  

The monthly composite files with extensions "avg\_rade9" contain floating point radiance values with units in nWcm$^{-2}$sr$^{-1}$ with the original DNB radiance values multiplied by one billion (1E9). The files with extension “cvg” are integer counts of the number of coverages or total observations available for that specific pixel that month; these were used to calculate the weighted three month averages.  

All spatial data were geo-referenced to an equal area projection, NAD 83 Albers with metric units. The VIIRS images were clipped to Chelan County boundaries and potential noise was removed by setting all values less than or equal to 0.5 to 0.0 \citep{duri18}, or in other words, we changed radiance values of 0.000000005 to 0.00.

We extracted two temporal datasets: one set contained the monthly composite clipped images for May 2018 and for May 2019, and the other set was a three-month average of April, May, and June for 2018 and for 2019. The vegetation coverage and the ground albedo were very similar during the study timespan because data were compared during the same months in both years. In May, our field trips confirmed that the city streets are completely free of snow. By using the average over the whole month of May and over the three-month period, we reduced the selection bias caused by the daily atmosphere variation.      

\subsubsection{Satellite Data Analysis}
Our satellite data analysis consisted of pre-processing the VIIRS imagery for use in the study, extracting (using zonal statistics) the radiance values within city and county boundaries (Stage 1), extracting streetlight data for comparison statistics (Stage 2), and image subtraction to visualize the differences (Stage 3). 

\subsubsection{Data Processing}
Stage 1. Zonal Statistics were used to get the sum of lights (SOL) within county and city boundaries. Zonal statistics are calculated for the pixels whose centers are located within the zone boundary. This tool runs descriptive statistics for pixel values (i.e., VIIRS radiance) within polygon boundaries, bridging the vector and raster data formats used in spatial analysis. Since the ground-based measurements were taken in May, we extracted the SOL from May 2018 and May 2019 VIIRS composites within the city boundaries, which provided a direct month-to-month comparison with ground measurements. We also extracted SOL for the three-month averages in 2018 and 2019 to reduce the monthly variance in VIIRS measurements \citep{coes18}.

Stage 2. We identified all of the pixels that had any streetlights present.  To extract the radiance per pixel, a fishnet dataset was created that aligned to the VIIRS imagery.  The following six attributes were created: 1) the VIIRS values for the month of May, 2) the VIIRS monthly average, 3) the number of streetlights in each pixel polygon, 4) binary-true-false attribute if all lights in pixel polygon were retrofitted (Complete), 5) binary-true-false attribute for a mix of retrofitted and not retrofitted lights (Partial), and 6) binary-true-false attribute for pixels with unmodified street lighting (None).  A paired Student t-test was run with these variables.  

Stage 3.  We subtracted the two temporal datasets:  May 2019 minus May 2018, and Average 2019 minus Average 2018.  Overlaying this information with the retrofitted and not retrofitted streetlights within each city, we examined the spatial information. Image differencing calculated the changes between pixels in two images on a pixel-by-pixel basis to give a quantitative and  visual assessment of temporal changes collected by the satellite data. 

\subsection{Camera Measurements of Sky Brightness}
\subsubsection{NPS Camera System and Data Collection}
We measured the brightness of the night sky by taking images with the NPS camera system. The system included a Nikon 50mm f/1.2 lens, a \textit{V}-band filter, and a research-grade charge-coupled device (CCD) \citep{duri07}. The system had $24^\circ$ x $24^\circ$ field of view with image resolution of 1.4 arcminute/pixel. A complete image acquisition sequence takes approximately 40 minutes and yields a data set of 45 images forming a hemispherical view. The broadband $V$ filter approximates the astronomical Johnson-Cousins \textit{V} filter and lets only visible light pass through, allowing the detected signal to closely represent what human eyes can see with photopic vision. There often exists some differences between sensor sensitivity, human vision, and light sources (eg. \citealt{sanc17}). At an urban transition site like this, the peak sensitivity of human’s nighttime vision is shifted slightly towards the blue. 

The data were collected on Burch Mountain (\ang{47;32;24}N, \ang{120;21;20}W), located approximately 13 km NNW of Wenatchee. This site was selected based on its relatively high elevation (1,250 m), which offered a clear view of the horizon above the largest population center in Chelan County. No other large population centers are along the SSE line-of-sight within 300 km, so the contamination from other distant lights was minimal. 

Two nights of data were collected: one in 2018, before the retrofit started; and one in 2019, after its completion. Details of each data collection event are listed in Table~\ref{table:events}. The extinction coefficients were measured from the collected images to characterize the attenuation of light due to absorption and scattering of the atmosphere. All the data were collected when the sun and the moon were more than 18$^\circ$ below the horizon.

\begin{table}
  \centering
  \begin{threeparttable}
    \caption[]{Data sets collected on Burch Mountain \label{table:events}}
    \begin{tabular}{lcccl}
      \toprule
      Date & Local Time (h) & Extinction Coefficient (mag/airmass) & Sky Condition\\
      \midrule
      2018-05-07 & 22.8 & $0.296 \pm 0.006$ & partly cloudy and humid\\
      2018-05-07 & 23.4 & $0.284 \pm 0.004$ & partly cloudy and humid\\
      2018-05-08* & 0.1 & $0.256 \pm 0.002$ & clear\\
      2019-05-31* & 23.6 & $0.266 \pm 0.002$ & clear but hazy during the day\\
      \bottomrule
    \end{tabular}
    *Best data set collected of the night.
  \end{threeparttable}
\end{table}

To complement the hemispherical image, visual assessments of night sky condition were recorded. Bortle Class was assigned based on visible celestial objects and appearance of the night sky on an integer scale from 1-9 \citep{bort01}. Limiting magnitude was determined by identifying the brightness of the faintest star visible with naked eye. Zenith brightness also was measured with the hand-held Unihedron SQM. The SQM has the half width at half maximum angular sensitivity $42^\circ$.

\subsubsection{Night Sky Image Processing}

The observational data from Burch Mountain were processed using the NPS pipeline \citep{duri07}. The pipeline performs basic data reduction including bias and dark subtraction, flat fielding, and detector linear response correction. Absolute brightness and position calibration were done by using the standard stars captured in the images. The calibrated images in a set were then stitched together to form panoramic images with the resolution of 0.05 degrees per pixel in \textit{V} band.

The extinction coefficient for each data set was measured to characterize the atmospheric conditions. If the atmospheric aerosol content is high, the extinction coefficient will be high because it diminishes the star light via increased absorption and scattering. Thus, measuring extinction coefficients is a way to probe the optical property of the atmosphere. The coefficient was measured using all the standard stars detected in the images. On average, about 125 stars were used. They spread across the sky in all azimuthal directions and from zenith to about 12 degrees above the horizon. We measured the brightness of those stars and compared them to their absolute brightness. By plotting the airmass derived based on their altitudes and the difference of their absolute and apparent magnitudes, the points formed a linear trend. We measured the extinction coefficient by finding the slope through the least square fit. Table~\ref{table:events} listed the extinction coefficient measured in each data set along with its uncertainty. The method we used here is the same as the one published in the original research \citep{duri07}. The measured extinction coefficients were very close to each other, which means that the atmospheric conditions, presumably the aerosol content, were very similar.    

To obtain some sky brightness metrics with anthropogenic light only, we subtracted out the natural sky brightness. A natural sky model \citep{duri13} was built based on the time and the location of the observation. Components of natural sky model included zodiacal light, Milky Way, airglow, and atmospheric diffuse light (ADL). Airglow and ADL were assumed to be azimuthally symmetric. The model parameters were set to the same standard values across the data sets except actual measured atmospheric extinction coefficients were used for the zodiacal light, Milky Way, and airglow components. The ADL accounts for the general brightening of the sky near horizon because of multiple scattering, and the extinction profile for the ADL was adopted directly from \citet{duri13}.

\section{Results}

\subsection{Changes in Upward Radiance}
In Stage 1, once we extracted the mean radiance values and the sum of the lights (SOL) within the county and city boundaries of Chelan County. The difference in SOL for Chelan County from both temporal data sets was negative. This indicates that the entire county was darker in 2019 than in 2018 as detected by the VIIRS sensor. Then we did the same for the five cities. The mean radiance values for the cities from May 2018 to May 2019 extracted from the May VIIRS composites overall went down. The differences between SOL values for the month of May in 2019 and 2018 showed all five cities had a lower SOL value in 2019 (Table~\ref{table:subDiff}).  The 2019 US Census of the city populations were provided in the table as a measure of the city size. For the three-month average data set, all five cities had a negative mean difference (Table~\ref{table:subDiff}). This indicates less detectable light in 2019. From the three-month average images, again all five cities had negative SOL values, lower in 2019 than in 2018. Recall a negative difference indicates that there was less light detected in 2019 than in 2018 in Chelan County. 

\begin{table}
\centering
  \begin{threeparttable}
    \caption{Difference between Mean Radiance and Sum of Lights (SOL) for Cities in Chelan \label{table:subDiff}} 
    \begin{tabular}{lcrrrrrrr}
      \toprule
       City   &  Population & Pixel  & \multicolumn{2}{c}{May Difference} & \multicolumn{2}{c}{Three month Average Diff}\\
       Name & 2019 & CNT & MEAN & SOL & Mean &  SOL \\
      \midrule
       Cashmere    &3,172 & 15          & -3.601      & -54.020   & -3.367         & -50.508    \\
       Chelan      & 4,237 & 81          & -0.351      & -28.420    & -0.020         & -1.636     \\
       Entiat     &1,280  & 26          & -0.381      & -9.920    & -0.576         & -14.984    \\
       Leavenworth &2,029 & 17          & -0.021      & -0.349     & -0.057          & -0.968     \\
       Wenatchee   & 34,360 & 123         & -1.957      & -240.670   & -2.891         & -355.549    \\
      \bottomrule
    \end{tabular}
  \end{threeparttable}
\end{table}

Stage 2 of our analysis created a "fishnet" mesh of polygon boundaries aligned with the VIIRS imagery pixels to create 731 pixel polygons. We then counted the streetlights within each pixel polygon and separated the polygons into three groups: Complete: all lights in the pixel polygon were retrofitted (127); Partial: the pixel polygons included a mix of retrofitted and not retrofitted lights (258); and None: no changes or retrofits in street lighting (346). Running a paired Student t-test revealed that the VIIRS values were significantly lower in 2019 for both the Complete and Partial groups, and not significantly different for None group  (Table~\ref{table:ttest}).

\begin{table}
\centering
  \begin{threeparttable}
    \caption{t-test of the VIIRS radiance values from 2018 and 2019 for pixels with streetlights \label{table:ttest}} 
    \begin{tabular}{lccccc}
      \toprule
       Streetlights & Number of  & \multicolumn{2}{c}{May} & \multicolumn{2}{c}{Three month Average}\\
      Retrofits & Pixel Polygons & t-stat & p-value & t-stat & p-value \\
      \midrule
       Complete & 127 & 2.592 & 0.010 & 2.862 & 0.005 \\
       Partial & 258 &  5.826 & 0.000 & 8.701 & 0.000 \\ None & 346 & -1.600 & 0.110 & -1.592 & 0.113 \\
      \bottomrule
    \end{tabular}
  \end{threeparttable}
\end{table}

\begin{figure}
    \centering
    \includegraphics[width=0.6\textwidth]{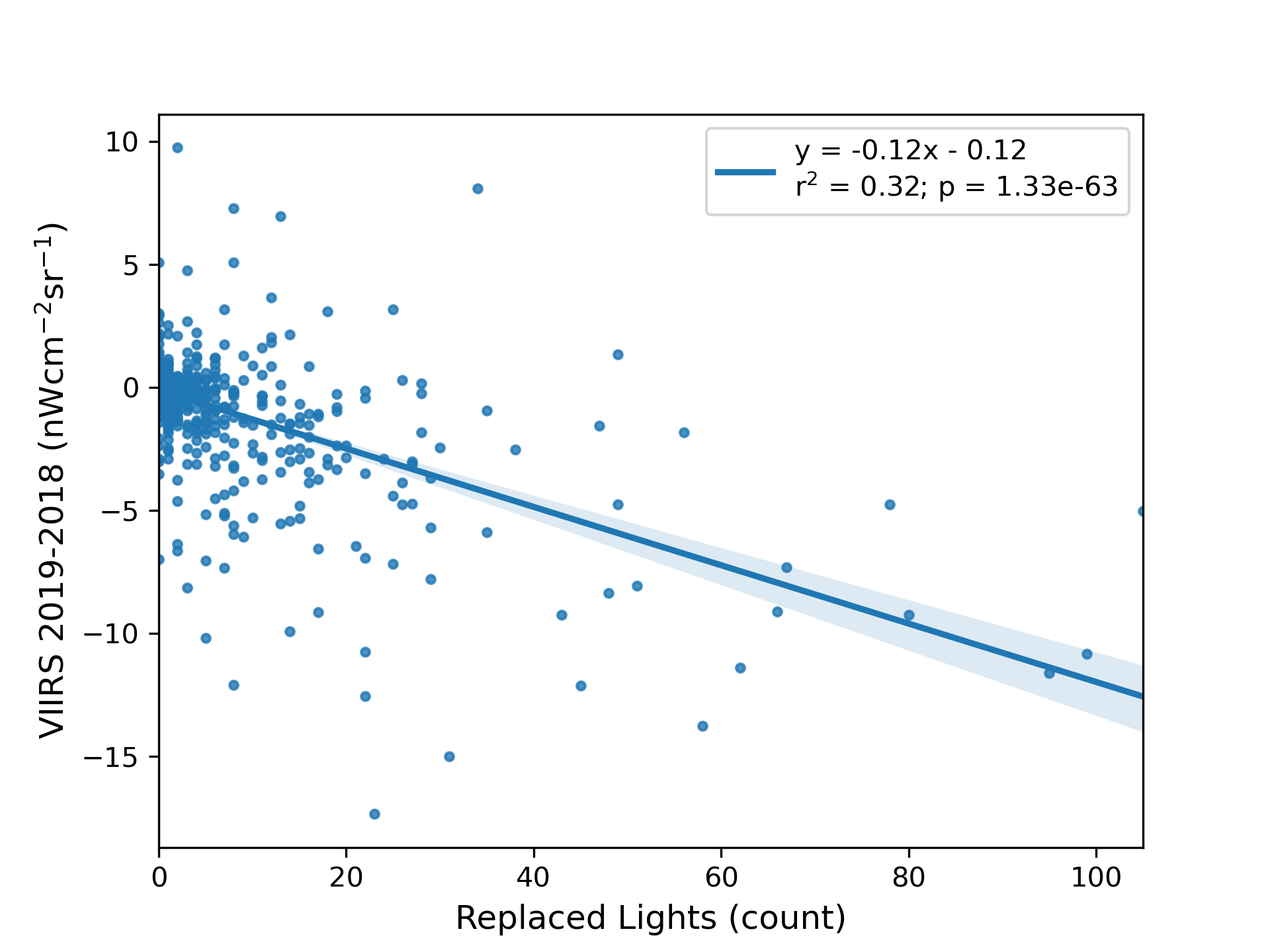}
    \caption{Regression plot showing a decrease in VIIRS radiance values as the number of lights are retrofitted increases. The shaded area show the 68\% confidence interval.}
    \label{SD_regressPlot}
\end{figure}

To understand if there was a significant decrease with retrofitted lights, we ran a simple linear regression of the retrofitted lights and the difference between the VIIRS radiance values. This regression represents only those pixels that had lights replaced. The independent variable (Count Replaced) is the number of lights that were replaced within that pixel (each point represents a pixel polygon). The dependent variable (change between 2019 \& 2018 or ‘delta’) represents the average radiance in that pixel in 2019 minus the average radiance in that pixel in 2018. The linear regression showed that when the number of replaced bulbs increases, delta decreases, demonstrating that the VIIRS satellite data reports less (or darker) average radiation as more bulbs are replaced with LEDs (Figure~\ref{SD_regressPlot}).

\begin{figure}
    \centering
    \includegraphics[width=0.9\textwidth]{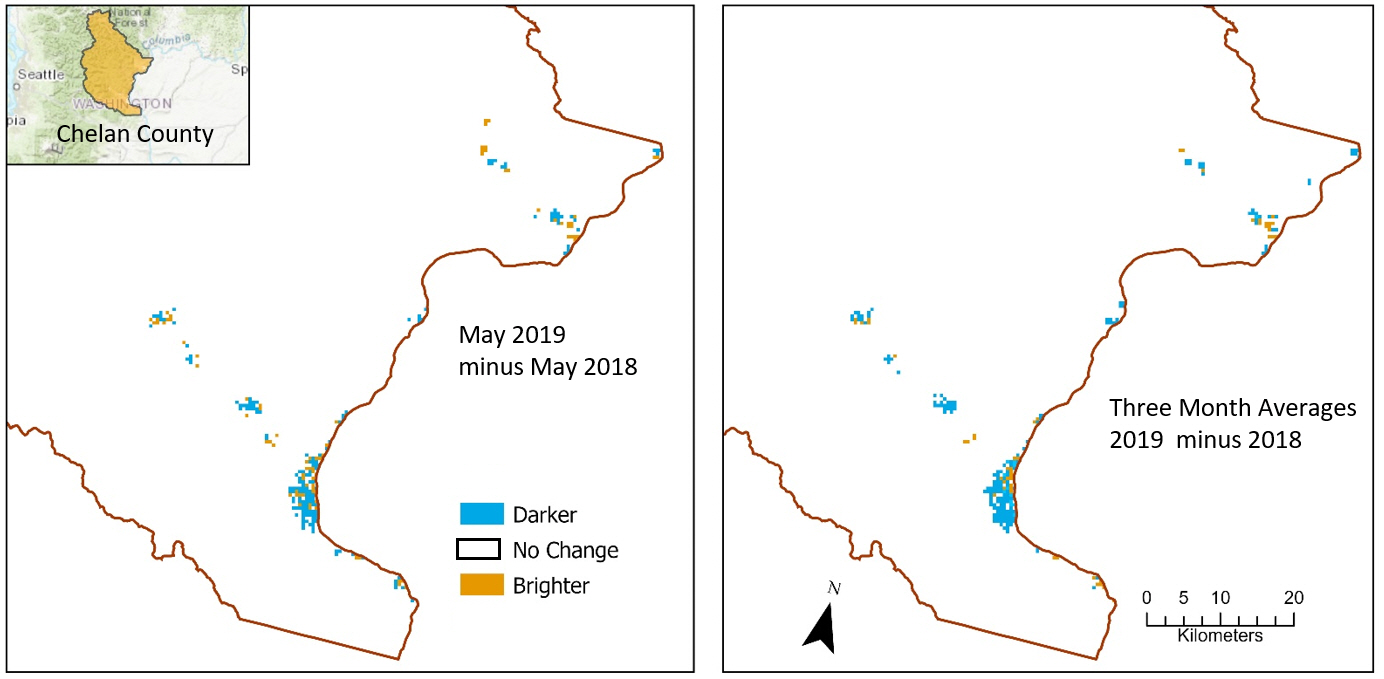}
    \caption{Difference between VIIRS 2019 and 2018 for the month of May and for the 3 Month Average of April, May, and June.}
    \label{SR_ChlnChange}
\end{figure}

Stage 3 was a simple image subtraction of the VIIRS data to visualize the differences between the before and after lighting retrofit: the temporal 2019 datasets  minus the 2018 datasets  (Figure~\ref{SR_ChlnChange}). The areas of blue are darker in 2019; white is the same both years; and gold is brighter in 2019. Combining the information with the retrofitted and not retrofitted streetlights within each city yields a spatial display of the information. The visual assessment of these data supports the statistical analysis. The majority of Chelan County remained the same, since most of the county is undeveloped and without lighting. The overall reduction in radiance identified by VIIRS data mainly occurred in the cities where streetlights had been retrofitted.  

Of the 3693 retrofitted streetlights in Chelan County, 3250 are completely within the boundaries of the five cities. A closer look at the impact of the retrofit on the cities shows an overall decrease in radiance received by the VIIRS sensor for all five of them  (\nameref{appendixa} contains large scale maps for each of the cities within Chelan County).  

\subsection{Changes in Sky Brightness}
\begin{figure}
    \centering
    \includegraphics[width=\textwidth]{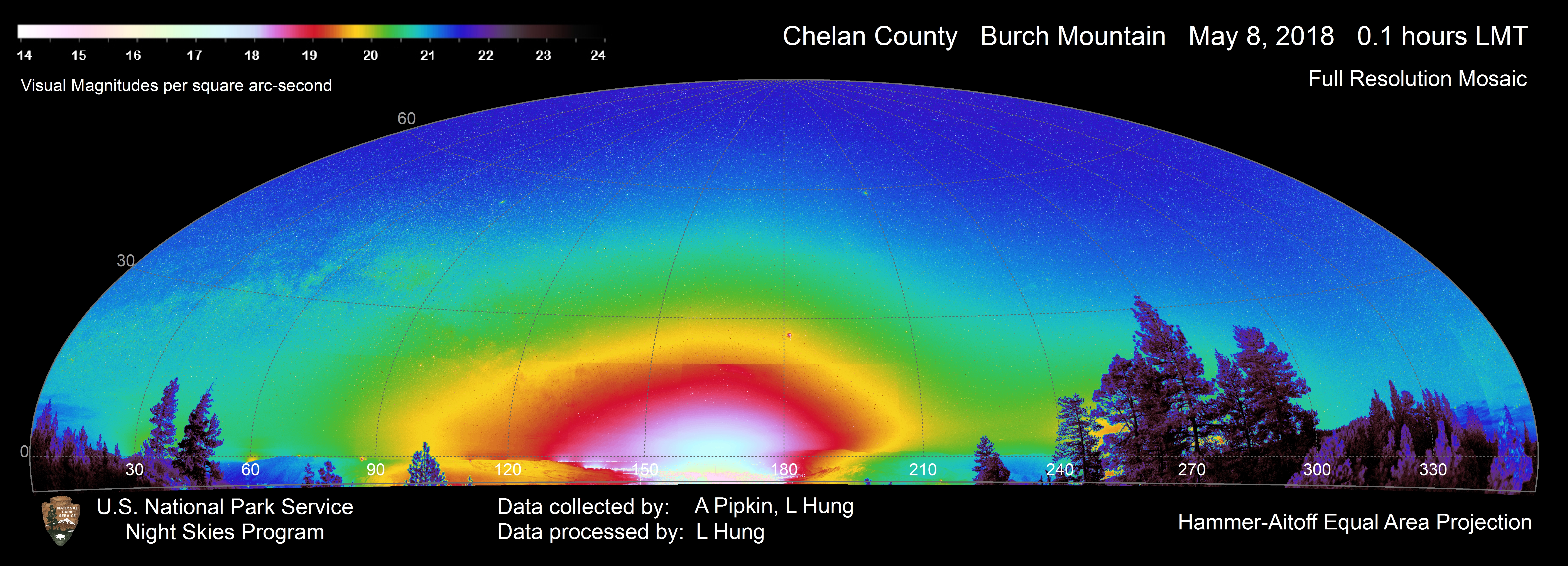}
    \includegraphics[width=\textwidth]{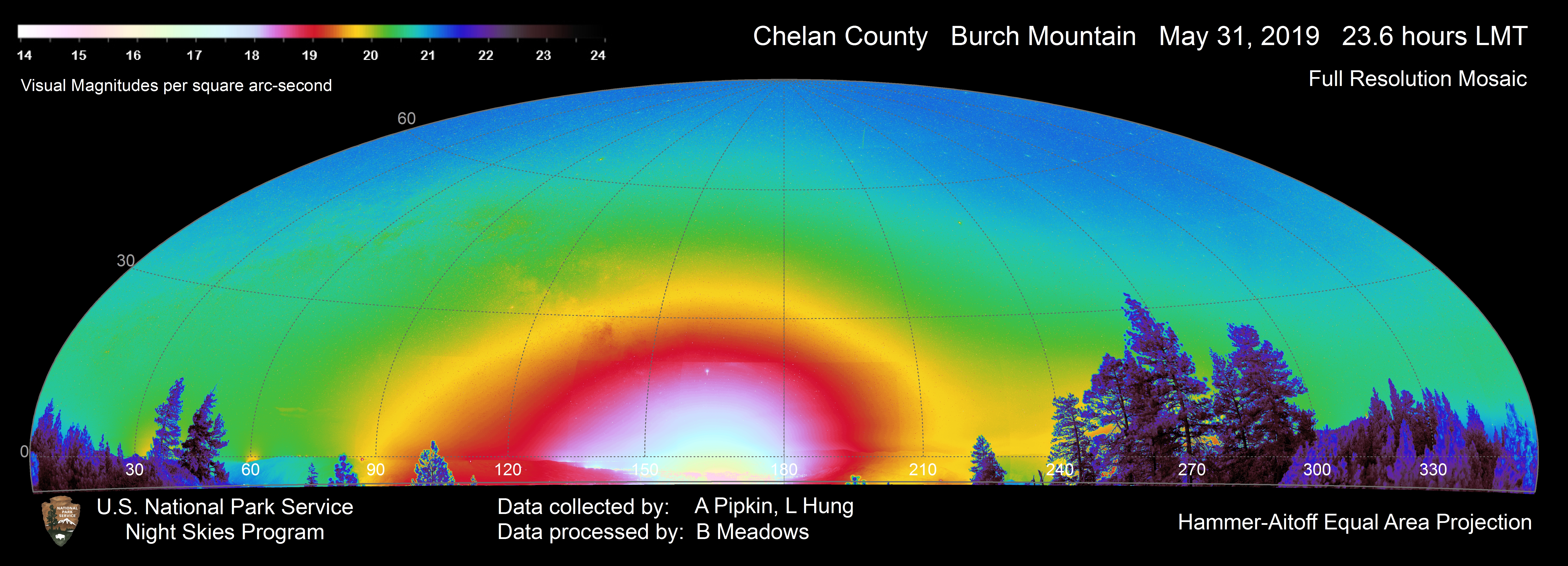}
    \caption{Panoramic \textit{V}-band images taken before (top) and after (bottom) the lighting retrofit on Burch Mountain. These observed images capture light from both natural and artificial sources. Sky brightness is displayed with the false colors in mag/arcsec$^2$. The observations show that the light dome became bigger and extended higher into the sky. Overall, the sky became brighter after the LED lighting retrofit.}
    \label{panoramicmosaic2018}
\end{figure}

The observations on Burch Mountain show that the sky became brighter after the lighting retrofit. Figure~\ref{panoramicmosaic2018} shows the calibrated \textit{V}-band images taken in 2018 (top) and 2019 (bottom), corresponding to before and after the lighting retrofit. South is centered in the images. The Milky Way appears at the similar sky location in both years, reaching about $45^\circ$ above the eastern horizon. The big light dome near the image center is from Wenatchee, the largest population center in Chelan County. The colorbar shows the level of sky brightness measured in magnitude per square arcsecond (mag/arcsec$^2$). 

A sky background brightness of 22 mag/arcsec$^2$ is considered pristine \citep{gars89, pata08}, and a sky background brightness $<20$ mag/arcsec$^2$ would be considered substantially altered from the natural condition. Purple and dark blue indicate unpolluted sky, and the Milky Way under the natural condition appears green in this color scheme. In 2019, most of the night sky has greatly deviated from the natural condition. Comparing the 2019 image to the 2018 image, the Wenatchee light dome appears larger and extends higher into the sky. The Milky Way is visible in both years but it fades into the city light dome at a higher elevation after the lighting retrofit. The zenith also shows skyglow.

We quantify this change with the sky brightness metrics derived from the images. Five indicators \citep{duri16} based on the observed panoramic images focus on different aspects of sky brightness: horizontal illuminance, maximum vertical illuminance, zenith brightness, percentage of lost stars, and all-sky light pollution ratio (ALR). Illuminance, the amount of visible light incident on a unit surface area, quantifies how brightly the landscape is lit from the night sky. The zenith brightness is the sky brightness overhead. The lost star metric estimates the percentage of stars that become invisible under the influence of skyglow. ALR is an index of total skyglow calculated by dividing the total brightness of the skyglow by the brightness of the natural dark sky. These quantitative measurements are summarized in Table~\ref{table:indicators}. 

\begin{table}
\centering
  \begin{threeparttable}
    \caption{Sky Brightness Metrics \label{table:indicators}}
    \begin{tabular}{lccc}
      \toprule
       & 2018-05-08 & 2019-05-31 & Natural Reference\\
      \midrule
      Horizontal Illuminance (mlx) & 1.55 & 2.31 & 0.80 \\
      Max. Vertical Illuminance (mlx) & 3.85 & 5.23 & 0.40 \\
      Zenith Brightness (mag/arcsec$^2$) & 21.60 & 21.10 & 22.00 \\
      Lost Stars$^\dagger$ (\%) & 39 & 51 & 0 \\
      All-sky Light Pollution Ratio$^\dagger$ & 2.30 & 3.69 & 0 \\
      Bortle Class & 6 & 5 & 1 \\
      Limiting Magnitude & 5.6 & 5.8 & 7.0 \\
      SQM (mag/arcsec$^2$) & 21.20 & 21.03 & 22.00 \\
      \bottomrule
    \end{tabular}
    $^\dagger$metrics derived with the natural sky brightness modeling
  \end{threeparttable}
\end{table}

Two visual assessment metrics are also reported here: Bortle Class and limiting magnitude. Bortle Class 6 corresponds to "bright suburban sky" whereas 5 corresponds to "suburban sky." Limiting magnitude of 5.6 and 5.8 indicate the brightened sky background limits the faintest star visible with naked eye. The final metric is a measure of zenith brightness determined with the hand-held Unihedron SQM. The natural sky reference value of each metric is listed in the last column of Table~\ref{table:indicators}.   

Overall, the night sky on Burch Mountain resembles suburban sky that is getting brighter. The decrease of 0.5 mag/arcsec$^2$ for zenith brightness indicates the sky overhead on Burch Mountain was 60\% brighter in 2019 than in 2018. The visual impact of the skyglow includes losing an estimated 51\% of the stars visible to the naked eye. When considering the entire sky, the ALR values indicate the total artificial light was 2.30 times the natural light level in 2018 and increased to 3.69 in 2019. The Bortle class and limiting magnitude are assessed visually, in this case by the same two researchers on-site. Visual assessment is more variable than the other methods of data collection, and the changes measured here were smaller than the human eye can confidently perceive, which means that the reported values for these metrics are not as reliable as those of the other metrics. In summary, based on our calibrated and objective measurements, the sky was brighter in 2019, after the LED retrofit.

\section{Discussion \& Conclusion}

The VIIRS satellite did not detect the increase of skyglow due to the LED lighting retrofit. Our results show that the brightness of the retrofitted area decreased from the satellite's view, even though the measured skyglow increased. These decreases in VIIRS radiance measurements were probably due to improved directivity, shielding, and shift in spectrum of the LED streetlights. Our findings support previous claims \citep{kyba17} that VIIRS DNB, due to its “blue blindness,” may falsely suggest a reduction in light pollution in many cities after an LED retrofit. 

Using satellite imagery has become an increasingly popular and accessible way to estimate skyglow around the globe. For example, Light pollution map \citep{star19} and NASA's Black Marble \citep{roma19} offer simple web interfaces for public use. However, prior models relating VIIRS DNB data to skyglow will – at a minimum – require substantial revision to account for the different characteristics of solid state luminaires. This research is critical for the local communities, lighting industry and scientific community in understanding how lighting retrofits affect satellite measurements and modeling of light pollution. 

We feel confident that our ground observations reflect the actual sky brightness change due to the lighting retrofit because those observations were taken under similar conditions. Our finding that the sky get brighter is based on the two nights of measurements. With a small sample size, selection bias needs to be considered. Natural sources of brightness can vary from night to night and sometimes in shorter time scale like in minutes or hours. Although the natural variation could affect the measurements, the designed observation time allow us to eliminate some seasonal and hourly variations from sources such as the Milky Way and the zodiacal light. Furthermore, the measured extinction coefficients show that the atmosphere exhibits similar scattering and absorption property on those two nights. Because of the similarity of the observing time and condition, we report our findings directly from the observed images. 

Comparing the images before and after the retrofit, the light dome over Wenatchee became brighter and extended higher into the sky. The zenith sky brightness on Burch Mountain increased by 60\%. The actual brightening based on human night vision is likely to be more than we measured. LEDs emit a signature blue peak at about 450 nm, which does not pass through the V-band filter attached to our camera. If the light from the blue peak were captured, we would expect the results to show the sky had brightened even more.

Though expert visual assessments in the field suggested the sky background became darker, our objective measurements showed otherwise. This reminds us that visual sky assessments, even by professionals, are not reliable for assessing the effects of lighting upgrades. Sky brightness is better measured by radiometric methods. Calibrated camera images are the most accurate method to document changes in the radiance and shape of light domes.

This case study shows that typical one-for-one HPS to 3000K LED replacements are likely to increase light pollution. Chelan County retrofitted all of their 3,600+ county-owned street lights in a period of one year. The county staff intentionally chose mostly 3000K LEDs, a color temperature generally regarded as night sky friendly. Furthermore, the estimated average lumen output per light decreased by about 50\%. The new LEDs are more energy efficient, have lower brightness, provide better directed light beams, and are fully shielded. Yet, these advantages did not result in reduced skyglow.

The main motivations for streetlight upgrades are reduced costs due to lower energy use and longer lamp lifetimes. Unarguably, LEDs are more energy efficient than the old HPS streetlights. According to the data provided by Chelan County, the total energy savings achieved by the end of 2019 from this streetlight retrofit project topped 2.6 million kilowatt-hours --- enough energy to power 120 all-electric homes in the State of Washington for a year. Going forward, it is easy to see how the cost savings can add up rapidly over time, especially with the long expected lifetime of LED lights. Because of these benefits, many cities, including Chicago and Denver, are in the middle of large-scale LED retrofits, and many more cities are planning them. Understanding societal and ecological impacts from lighting retrofits is an increasingly important global concern. 

How do we convert to LED technology in a way that reduces energy consumption, reduces skyglow, preserves historical ambiance, and still meets the safety and security needs of our communities?

We believe with proper design and usage, energy savings and light pollution reductions can stand on the same side, both benefiting from LED retrofits. This study implies reductions in skyglow will require lowering overall light level (> 50\%), using warmer lights (< 3000 K), or both. Perceived conflicts between streetlighting objectives and light pollution can be further mitigated through controls that dim lights during periods of low traffic. More case studies are needed to pinpoint the exact criteria for a lighting retrofit to reduce skyglow. Nonetheless, a strong proactive lighting design is the key to converting to LED lights without increasing light pollution.

\section*{Acknowledgements}
The authors thank Jim White from Chelan County for providing lighting retrofit data, including luminaire specification and energy savings; Bob Meadows for processing data and selecting observation sites; and Kathlyn Nuessly for assisting with graphic design. 

\appendix
\section*{Appendix A: Five cities in Chelan County, Washington \label{appendixa}}

This appendix provides a detailed description of the changes within the five main cities in Chelan County. Table ~\ref{table:strLights} summarizes the retrofitted lights in the cities. The number of retrofitted lights within the five cities in Chelan County varied from 107 to 3390. Table~\ref{table:monthSOL} shows that all of the cities had a decrease in the mean value and the SOL for both the May-to-May and 3-month average data sets. This indicates less light was sensed by the satellite. The mean radiance values for the cities from May 2018 to May 2019 extracted from the May VIIRS composites overall went down. The differences between SOL values for the month of May in 2018 and 2019 showed all five cities had a lower SOL value in 2019 (Table~\ref{table:monthSOL}). The population of the cities were provided in the table as a proxy for the size of the cities. The percentage of retrofitted lights in each city ranges from 60\% to 82\%. This table is followed by the differences in the three month average data set, again showing that all five cities had a negative difference (Table~\ref{table:3avgSOL}).

\begin{table}
\centering
  \begin{threeparttable}
    \caption{Streetlights Replaced within  Cities in Chelan County \label{table:strLights}} 
    \begin{tabular}{lccrrr}
      \toprule
       CityName    &Streetlights & Replaced & Percent Retrofitted \\
       \midrule
       Cashmere    & 707          & 578      & 82      \\
       Chelan      & 321          & 157      & 48      \\
       Entiat      & 107          & 81      & 76       \\
       Leavenworth & 301          & 240      & 80       \\
       Wenatchee   & 3390         & 2018      & 60   \\
    \bottomrule
    \end{tabular}
  \end{threeparttable}
\end{table}

\begin{table}
\centering
  \begin{threeparttable}
    \caption{Monthly May Mean Radiance and Sum of Lights (SOL) in nWcm$^{-2}$sr$^{-1}$ for Cities in Chelan County\label{table:monthSOL}}
    \begin{tabular}{lcrrrrrrr}
      \toprule
       & Percent Lights &Population & Mean & Mean & Mean Diff & SOL & SOL & SOL Diff\\
       City& Retrofitted & 2019 & 2018 & 2019 & 2019-2018 & 2018 & 2019 & 2019-2018\\
      \midrule
       Cashmere & 82 &  3,172 & 17.63 & 14.03 & -3.60 & 264.46 & 210.44 & -54.02 \\
       Chelan & 48 & 4,237 & 5.57 & 5.22 & -0.35 & 450.84 & 422.42 & -28.42 \\
       Entiat & 76 & 1,280 & 3.56 & 3.18 & -0.38 & 92.59 & 82.67 & -9.92  \\
       Leavenworth & 80 & 2,029 & 10.77 & 10.75 &-0.02 & 183.12 & 182.77 & -0.35 \\
       Wenatchee & 60 & 34,360 & 29.79 & 27.83 & -1.96 & 3663.98 & 3423.31 & -240.67 \\
      \bottomrule
    \end{tabular}
  \end{threeparttable}
\end{table}

\begin{table}
  \begin{threeparttable}
    \caption{Changes in Three Month Average dataset \newline of Mean Radiance and Sum of Lights (SOL)\newline in nWcm$^{-2}$sr$^{-1}$ for Cities in Chelan County \label{table:3avgSOL}}
    \begin{tabular}{lcr}
      \toprule
       &  Mean Diff  & SOL Diff\\
       City & 2019-2018 & 2019-2018\\
      \midrule
       Cashmere &  -3.36 & -50.51 \\
       Chelan &  -0.20 & -1.64 \\
       Entiat &  -0.58 & -14.98 \\
       Leavenworth &  -0.06 & -0.97 \\
       Wenatchee &  -2.89 & -355.55 \\
      \bottomrule
    \end{tabular}
  \end{threeparttable}
\end{table}

\begin{figure}
    \centering
    \includegraphics[width=0.75\textwidth]{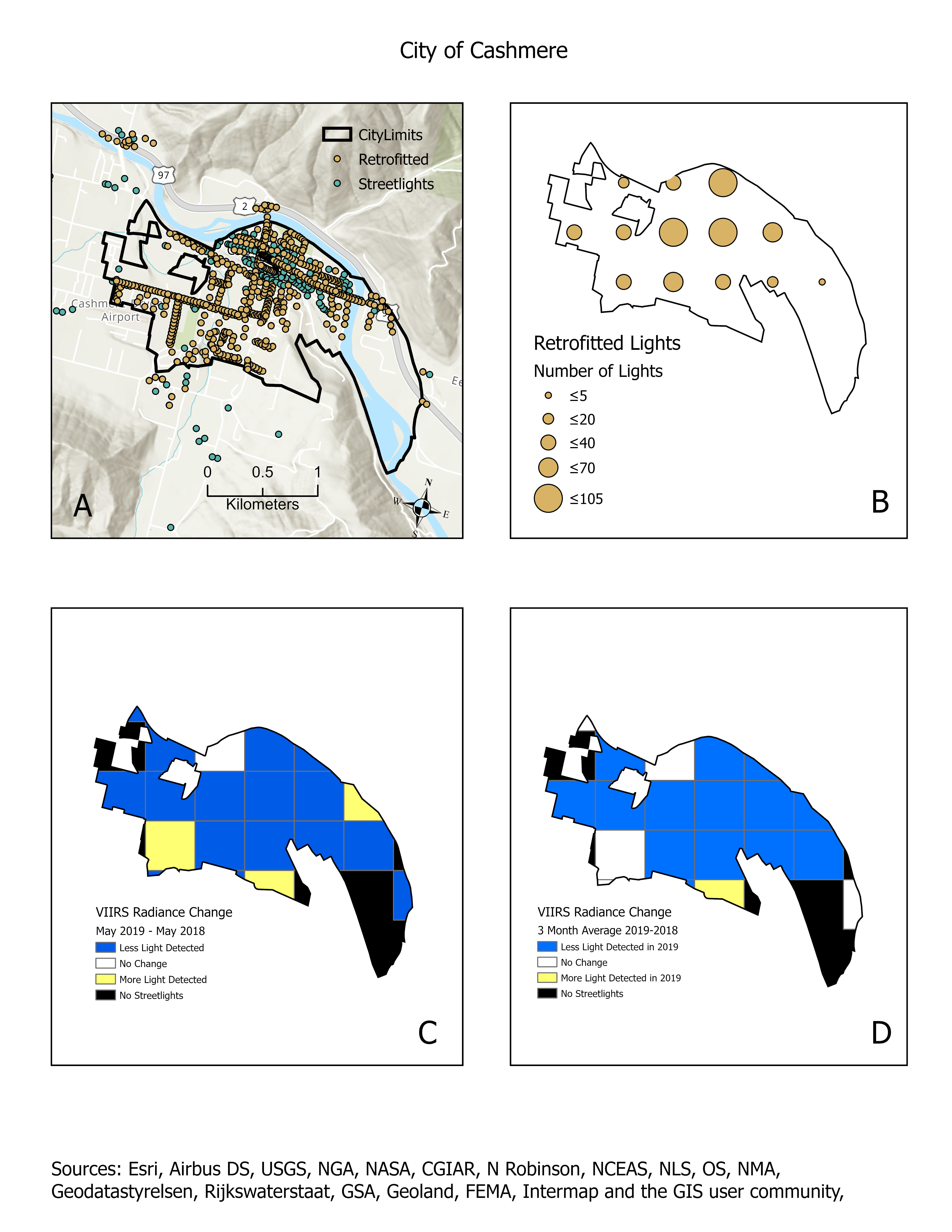}
    \caption{Example of Cashmere radiance changes between VIIRS 2019 and 2018 for the month of May and for the 3 Month Average of April, May, and June.}
    \label{A_Cashmere}
\end{figure}

\begin{figure}
    \centering
    \includegraphics[width=0.75\textwidth]{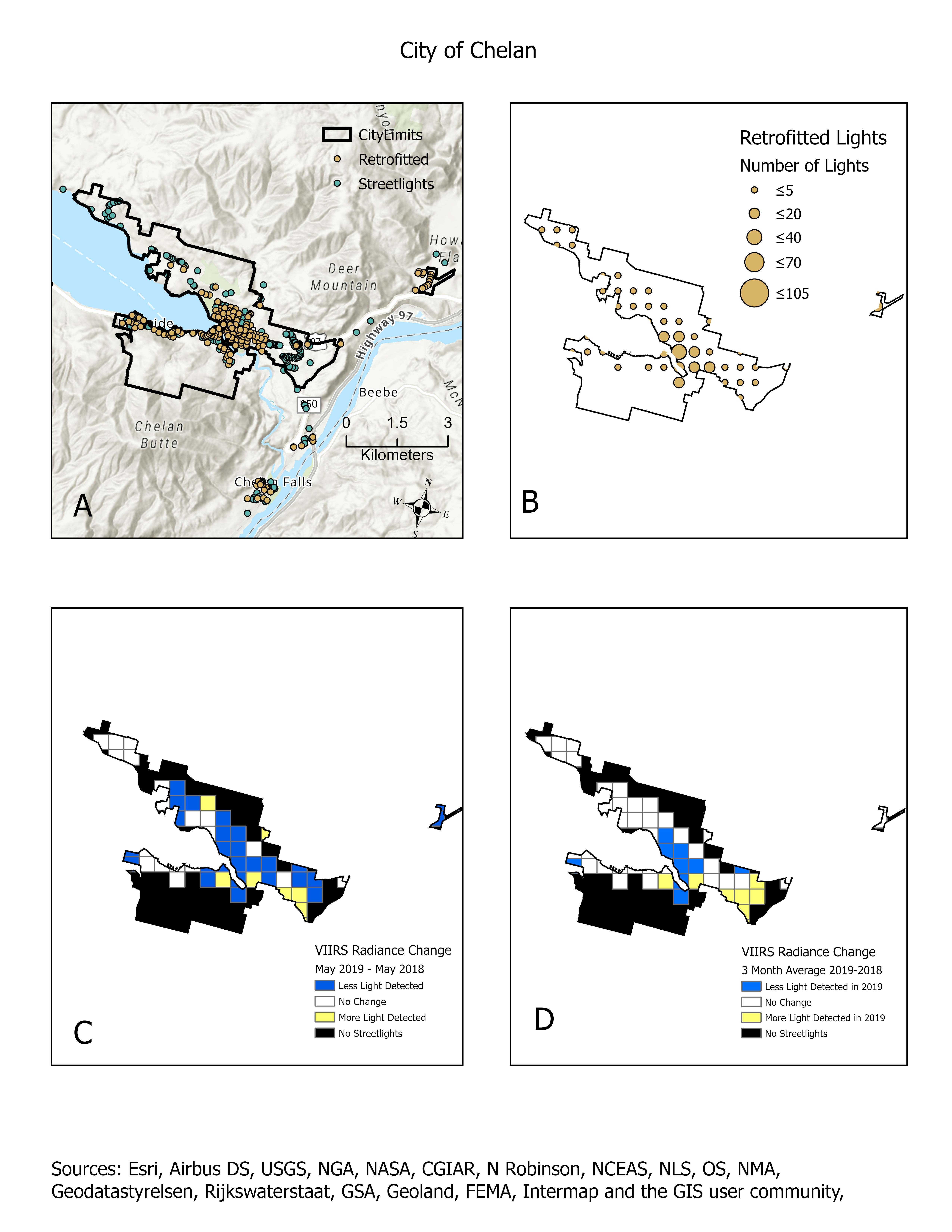}
    \caption{Example of Chelan radiance changes between VIIRS 2019 and 2018 for the month of May and for the 3 Month Average of April, May, and June.}
    \label{A_Chelan}
\end{figure}

\begin{figure}
    \centering
    \includegraphics[width=0.75\textwidth]{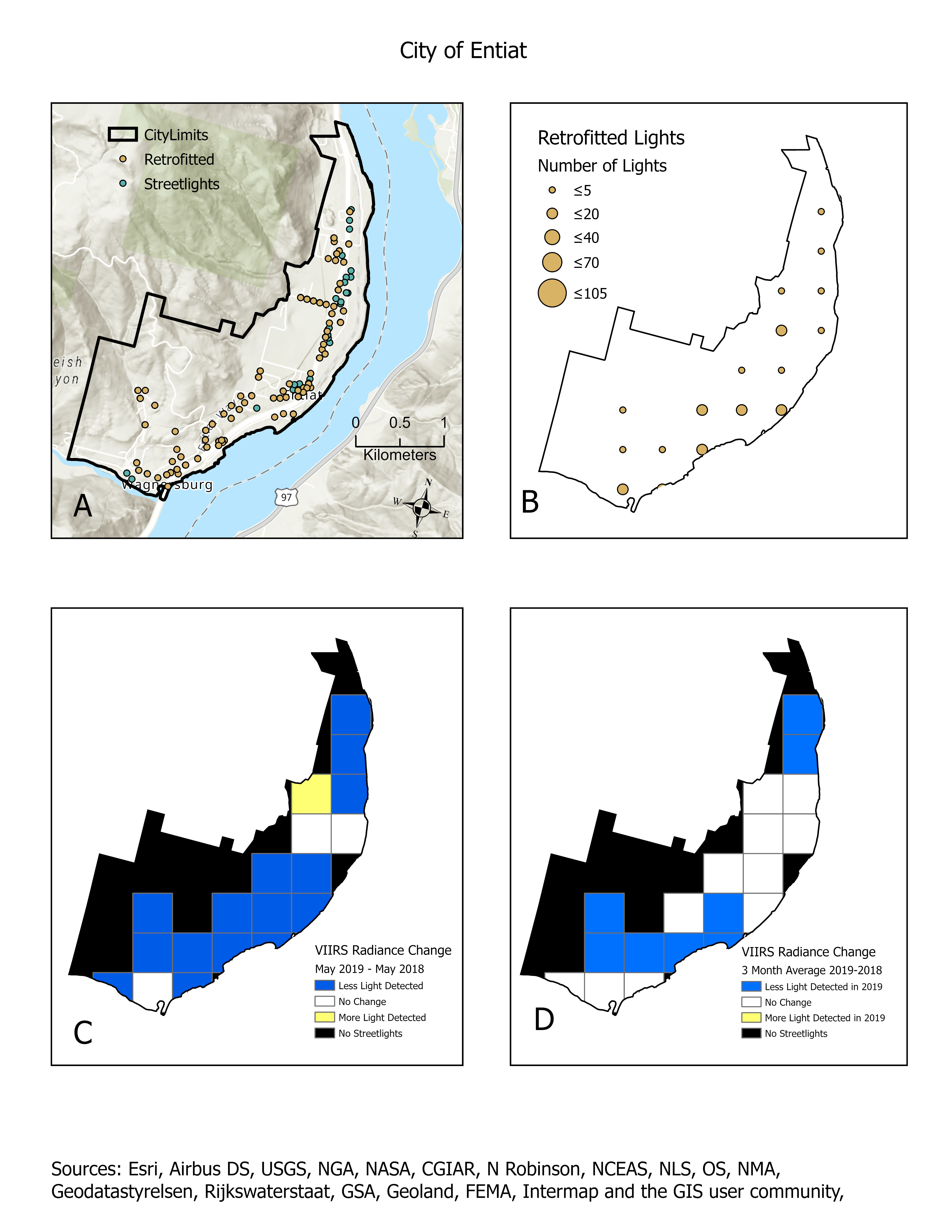}
    \caption{Example of Entiat radiance changes between VIIRS 2019 and 2018 for the month of May and for the 3 Month Average of April, May, and June.}
    \label{A_Entiat}
\end{figure}

\begin{figure}
    \centering
    \includegraphics[width=0.75\textwidth]{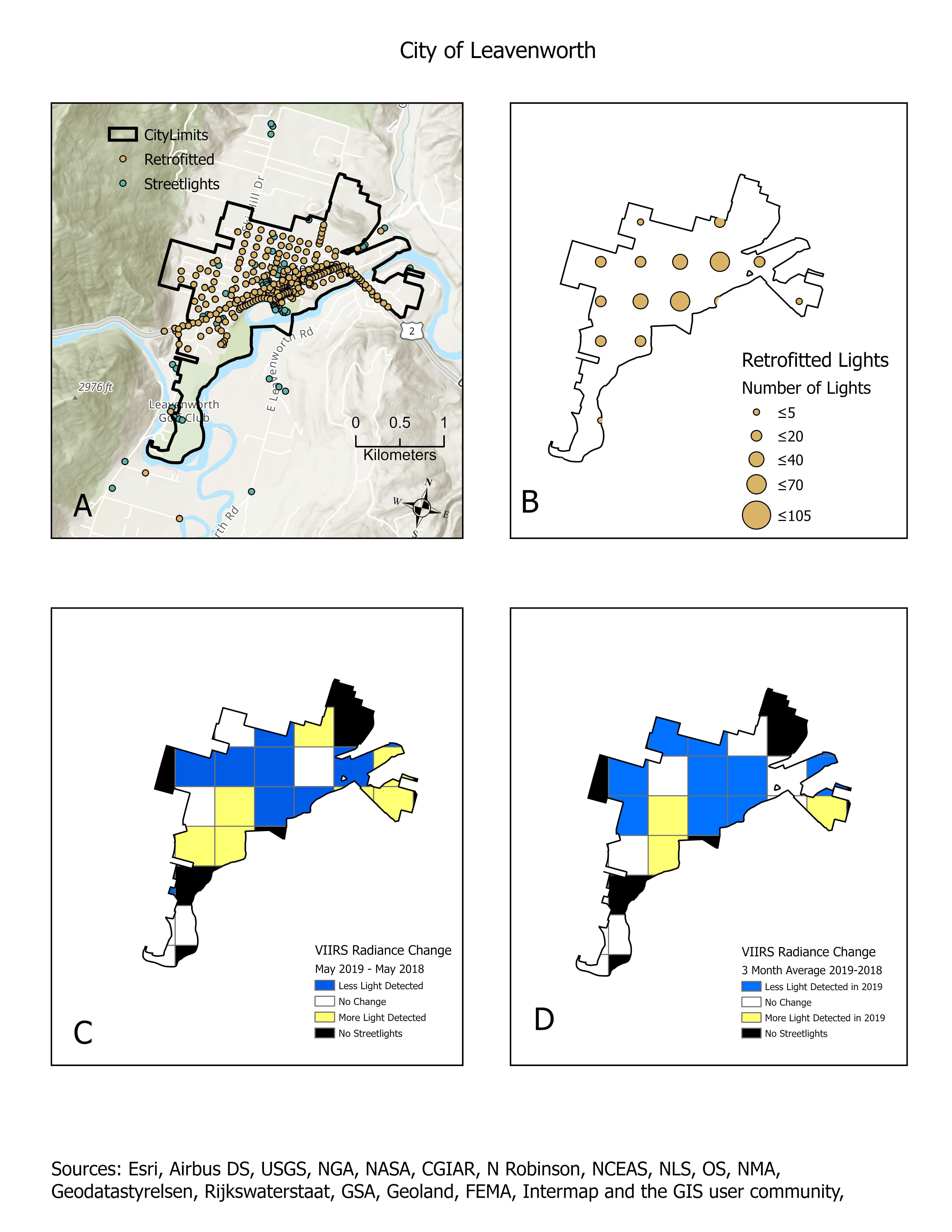}
    \caption{Example of Leavenworth radiance changes between VIIRS 2019 and 2018 for the month of May and for the 3 Month Average of April, May, and June.}
    \label{A_Leavenworth}
\end{figure}

\begin{figure}
    \centering
    \includegraphics[width=0.75\textwidth]{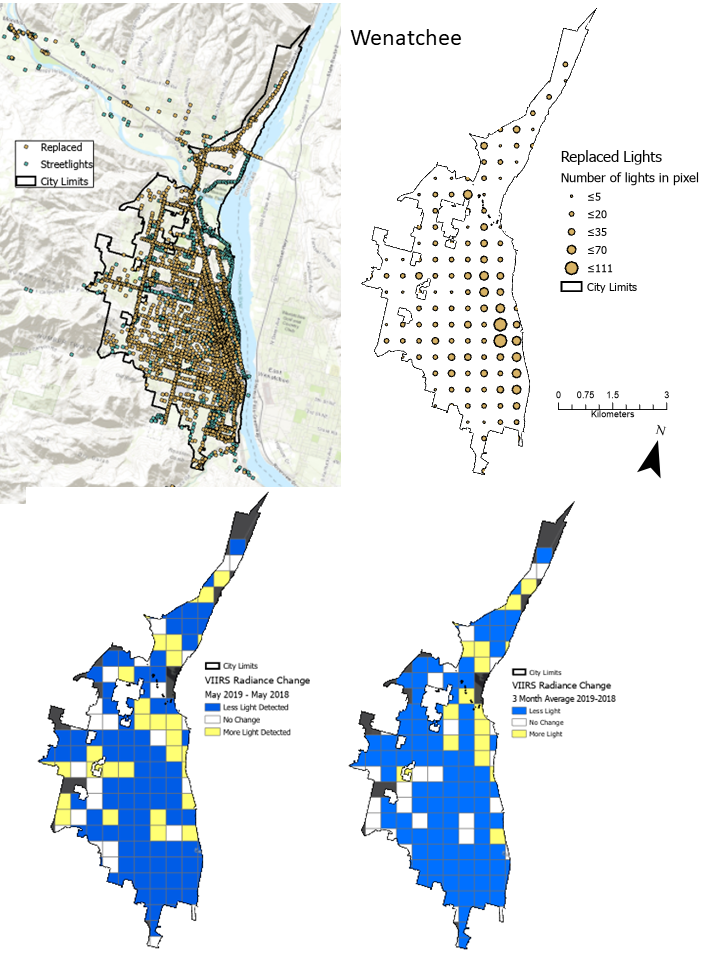}
    \caption{Example of Wenatchee radiance changes between VIIRS 2019 and 2018 for the month of May and for the 3 Month Average of April, May, and June. }
    \label{A_Wenatchee}
\end{figure}

Each of the five cities within Chelan county has four maps that provide geographical context about the retrofitted lighting within city limits: Cashmere (Figure~\ref{A_Cashmere}), Chelan (Figure~\ref{A_Chelan}), Entiat (Figure~\ref{A_Entiat}), Leavenworth (Figure~\ref{A_Leavenworth}), and Wenatchee (Figure~\ref{A_Wenatchee}).

Map A (top left) is a context map for cities. The base map has topography and water features with some landscape features labeled. All streetlights are shown with retrofitted (teal) and unchanged (khaki) lights. The scale bar applies to all four maps.

Map B (top right) uses graduated symbology to represent the number of lights retrofitted within the 500 m by 500 m pixel polygons.

Map C (bottom left) shows the change in upward radiance detected by VIIRS between May 2019 and May 2018. The negative values that had less light in 2019 are blue; if no change was detected, those cells are white; and the brighter values with more light after the retrofit are in yellow. This provides an overview of the May differences between the before and after retrofit measurements. 

Map D (bottom right) shows the change in upward radiance detected by the three-month average (April, May and June) of VIIRS monthly composites between 2018 and 2019. The negative values that had less light in 2019 are blue; if no change was detected, those cells are white; and the brighter values with more light after the retrofit are in yellow. Three month averages help mitigate the variance between VIIRS monthly composites.

\clearpage
\bibliographystyle{cas-model2-names}

\bibliography{references}

\end{document}